\documentclass[Royal,times,sagev]{sagej}

\usepackage{amsmath, amsthm, amscd, amsfonts, amssymb, graphicx, xcolor, float, algorithm, algorithmic, subcaption, booktabs}
\usepackage{tcolorbox}
\usepackage{multirow}
\usepackage{caption} 

\floatstyle{plain}
\newfloat{mybox}{tbp}{lob}
\floatname{mybox}{Box}
\restylefloat{mybox}

\theoremstyle{definition}

\theoremstyle{remark}

\usepackage[bookmarksnumbered, colorlinks, plainpages]{hyperref}

\title{Calculating the Expected Value of Sample Information accounting for missing data}

\author{
Luca Benetti$^{1}$,
Gianluca Baio$^{2}$,
Anna Heath$^{3,4}$\\[1em]
{\small $^{1}$Human Technopole, Milan, Italy}\\
{\small $^{2}$Department of Statistical Science, University College London, London, UK}\\
{\small $^{3}$The Hospital for Sick Children, Toronto, Canada}\\
{\small $^{4}$University of Toronto, Toronto, Canada}\\
}

\title{Calculating the Expected Value of Sample Information for Observational Studies affected by Confounding}

\begin{document}


\begin{abstract}
\textbf{Background}: The Expected Value of Sample Information (EVSI) quantifies the value of collecting additional evidence to inform a health economic model. Existing EVSI methods typically assume idealized data collection mechanisms, most commonly randomized controlled trials (RCTs). However, in many realistic contexts, additional evidence is expected to come from observational studies, which are subject to confounding or other bias. In this work, we define a methodology to calculate EVSI when the additional data are observational and potentially confounded. \\

\textbf{Methods}: First, we define a simulation-based framework in which observational data affected by confounding are generated through inverse target trial emulation (ITTE), a methodology that generates observational data with controlled levels of confounding starting from initial level data or prior information on population structure. Then, we apply inverse probability weighting (IPW) to obtain an adjusted summary statistic of the data targeting the corresponding randomized estimand. EVSI is finally computed using a regression-based approach. Moreover, we propose a computationally efficient method to determine the sample size required to recover the EVSI achievable under an idealized randomized design. \\

\textbf{Results}: We apply the methodology to two health economic models: a Normal–Normal conjugate model and a chemotherapy treatment model combining a decision tree and Markov structure. We show that EVSI computed from observational data, even after adjustment, is lower than EVSI based on randomized data. The loss in EVSI increases with the level of induced confounding. \\

\textbf{Conclusions}: This methodology extends EVSI to settings where future evidence is expected to be observational and affected by confounding, enabling value-of-information analysis in more realistic and feasible data collection scenarios. It also provides a principled approach to sample size planning when observational evidence is anticipated.
\end{abstract}

\Highlights{
1. We extend EVSI methodology to settings where additional evidence is observational and affected by confounding.\\
2. We show that EVSI decreases with increasing confounding level.
}



\maketitle

\section{Introduction}

The Expected Value of Sample Information (EVSI) \cite{ades2004expected} quantifies the expected economic benefit of collecting additional data to reduce parameter uncertainty in a decision model. Additional evidence can improve the precision of the model inputs, potentially altering the optimal decision under a cost-effectiveness framework. Estimation of the EVSI therefore provides a measure of the expected value associated with reducing decision uncertainty through further data collection.

EVSI has been widely used to determine whether further research is worthwhile \cite{kunst2019value, reeves2019three}. It prevents resources from being wasted on interventions that appear effective due to limited evidence and avoids unnecessary research on interventions already evaluated with sufficient precision. Moreover, the EVSI could be used to support study design by clarifying which data should be collected and how large a study should be to maximize value relative to cost.

Most existing EVSI applications rely on idealized data collection schemes, typically randomized controlled trials (RCTs) \cite{heath2022simulating, heath2024value}. Specific methods have been developed for particular outcomes and designs, such as survival data \cite{vervaart2022efficient} or cluster randomized trials \cite{welton2014expected}. However, these approaches assume that future evidence is generated under randomized allocation or similarly simplified mechanisms, e.g. when no missingness occurs in experimental data.

In many realistic settings, however, additional evidence is expected to come from observational studies rather than RCTs. Observational data are often more feasible and less costly, but they are typically affected by confounding, selection bias, and population imbalance \cite{rosenbaum2010design}. In such contexts, causal adjustment methods must be applied before inference and the naive summary statistic of the observational data does not generally target the same estimand as a randomized design.

Understanding how to compute EVSI when future evidence is observational is therefore crucial. It allows value-of-information analysis to be applied in settings where randomized studies are infeasible, and supports research prioritization under realistic data-generating mechanisms. In this paper, we introduce a novel methodology to compute EVSI when planning to collect observational data affected by confounding-induced population imbalance.

To achieve this, we combine EVSI methodology with inverse target trial emulation (ITTE) \cite{autore2026titolo}, which transforms randomized data into observational data exhibiting controlled imbalance. We then apply inverse probability weighting (IPW) to obtain an adjusted summary statistic targeting the randomized estimand. EVSI is computed using this adjusted statistic within the regression-based framework. Finally, we develop a computationally efficient method to determine the observational sample size required to recover the EVSI attainable under idealized randomized evidence.

This paper proceeds as follows. First, we define EVSI within the standard Bayesian decision framework. Second, we describe how to simulate observational data using ITTE and how to adjust for confounding using IPW. Third, we introduce the general framework to compute EVSI with observational data. Finally, we apply the methodology in two health-economic models and discuss the implications for study design and research prioritization.

\section{Methods}
\label{section 2}

\subsection{Expected Value of Sample Information}
EVSI computes the value of collecting data to reduce uncertainty in the parameters of an underlying health economic model. As such, consider a health economic model that is aiming to compare a set of $d=1,\ldots,D$ different interventions, each of which is valued using the net monetary benefit \cite{stinnett1998net} $\text{NB}_d(\boldsymbol{\theta})$, as a function of a set of relevant parameters $\boldsymbol{\theta}$. These parameters are typically estimated from data, meaning that they are subject to uncertainty. We define $p(\boldsymbol{\theta})$ as the joint distribution that represents the uncertainty associated with the set of parameters. In a standard Bayesian decision-making framework, to choose the best decision, one has to maximize $\mathbb{E}(\text{NB}_d(\boldsymbol{\theta}))$, where the expectation is taken both with respect to individual level variability and parameter uncertainty. The uncertainty due to $\boldsymbol{\theta}$ is typically evaluated by sampling $N$ values from $p(\boldsymbol{\theta})$ via Monte Carlo (MC) simulation \cite{metropolis1949monte}. The process of sampling $N$ values from $p(\boldsymbol{\theta})$ is known as either \textit{probabilistic sensitivity analysis} (PSA) or \textit{probabilistic analysis} (PA)\cite{heath2024value} .

Within this context, EVSI can be then defined as follows:
\[
\text{EVSI} = \mathbb{E}_{\boldsymbol{Y}}\left[ \max_{d} \mathbb{E}_{\boldsymbol{\theta} | \boldsymbol{Y}}\left( \text{NB}_d(\boldsymbol{\theta})\right) \right] - \max_{d} \mathbb{E}_{\boldsymbol{\theta}}\left( \text{NB}_d(\boldsymbol{\theta})\right),
\]
where $\boldsymbol{Y}$ represents the data we plan to collect to inform our decision model \cite{ades2004expected}. The outer expectation is taken with respect to the prior predictive distribution defined through $p(\boldsymbol{Y}\mid\boldsymbol{\theta})$ and $p(\boldsymbol{\theta})$ \cite{heath2022simulating}. In practice, EVSI is computed using efficient approximation methods, such as the regression-based method, moment matching or importance sampling \cite{heath2017review} \cite{heath2018efficient}. In this work, we compute EVSI using the regression-based method \cite{strong2015estimating}, which uses non-parametric regression to compute the EVSI. 

For the non-parametric approach to EVSI calculation, one first has to simulate a dataset $\boldsymbol{y}$ from the prior predictive distribution $\boldsymbol{y} \sim p(\boldsymbol{Y})$ by first sampling parameters $\boldsymbol{\theta}^* \sim p(\boldsymbol{\theta})$ and then $\boldsymbol{y} \sim p(\boldsymbol{Y}\mid\boldsymbol{\theta}^*)$. Once data are simulated, multiple nonparametric regression equations (one for each possible treatment) are fitted to compute the preposterior mean $\mathbb{E}_{\boldsymbol{\theta} | \boldsymbol{Y}}\left( \text{NB}_d(\boldsymbol{\theta})\right)$:

\[
\text{NB}_d(\boldsymbol{\theta}) =  g_d(T(\boldsymbol{y})) + \epsilon,
\]
where $\boldsymbol{\theta}$ is a sample of the joint distribution of the input parameters obtained by performing a probabilistic analysis (PA), and $T$ is some low-dimensional summary statistic of the data that summarizes the information in the data. Ideally, $T$ should include all the information in the original data; for this reason, it is suggested to choose sufficient statistics of the parameters we assume are informed by the data \cite{li2024estimating}. Other reliable options, when sufficient statistics are not obtainable, are typically sample estimators or maximum Likelihood estimators (MLE) of the parameter(s) we expect to inform collecting new data \cite{li2024estimating}. Under this approach $\mathbb{E}_{\boldsymbol{\theta} | \boldsymbol{Y}}\left( \text{NB}_d(\boldsymbol{\theta})\right) \approx \hat{g}_d\left(T(\boldsymbol{y})\right)$ and EVSI can be computed as:
\[
\text{EVSI} = \frac{1}{K}\sum_{i = 1}^K \max_d \hat{g_d}(T(\boldsymbol{y}^{(i)})) - \max_d  \frac{1}{K}\sum_{i = 1}^K \hat{g_d}(T(\boldsymbol{y}^{(i)})).  
\]

Current implementations of EVSI typically assume idealized study designs, where future data are generated under randomized treatment allocation and therefore directly target the estimand of interest without requiring causal adjustment. As a result, the assumed data-generating mechanism may not reflect how evidence is collected in many real-world settings. A methodology for computing the value of collecting data under realistic patterns of missingness was recently introduced \cite{benetti2026calculatingexpectedvaluesample}. In this work, we extend EVSI to settings where future evidence is expected to arise from observational studies affected by confounding. To achieve this, we first define a data-generating mechanism that simulates observational data with controlled levels of confounding.

\subsection{Simulating observational data with confounding}
\phantomsection
\label{simulate_full_data_obs}

To compute EVSI using the nonparametric regression approach, we must sample from $p(\boldsymbol{Y} \mid \boldsymbol{\theta})$. Here, $\boldsymbol{\theta}$ represents the parameter(s) from the health economic model that define the data generating process. The health economic model parameters $\boldsymbol{\theta}$ should represent the parameters estimated across the population of interest, i.e., the marginal estimand in the population. As such, these parameters are unbiased in the population of interest. This means that the initial data generating process will also be unbiased. For example, if estimating a treatment effect, the initial data generating process would be equivalent to data from an RCT. This RCT provides an experiment where the outcome distribution is coherent with the model parameters. From this idealized data set, we then generate observational datasets by inducing controlled confounding. 

Thus, our method for EVSI calculation begins by generating individual-level data from a hypothetical RCT (or similar idealized study) to preserve the relationship between the data generating mechanism and the true parameter values. We then introduce confounding by extending the data-generating process so that the conditional distribution also depends on individual-level covariates that are associated with both $\boldsymbol{Y}$ and the exposure variable $Z$. This approach allows us to preserve the connection between the health economic parameter(s) $\boldsymbol{\theta}$ and the resulting biased observational data while explicitly modeling the conditional distribution $Y \mid \boldsymbol{X}, Z$. In the next section, we describe how this extended framework is incorporated into the EVSI calculation.

\subsubsection{Modeling the complete randomized controlled trial}

To generate observational data with realistic confounding, we need to first specify a model to generate a complete dataset from a hypothetical individual-level synthetic RCT  \cite{heath2022simulating}. In many cases, the parameters governing the sampling distribution of these data will not be available in the health economic model and must be derived from the literature or expert knowledge. If the study team has access to data from an RCT, this information can also be leveraged to generate future study data.

In general, we assume that our proposed RCT will collect information from $n$ individuals. This information will include $S$ covariates $\boldsymbol{X} = \{X_1, X_2,.., X_S\}$, the exposure variable $Z$ and the outcome variable $Y$. To model the data from the hypothetical RCT, we can factorize the joint distribution of the covariates and the outcome variable into a product of sequential conditional distributions. In particular, for $X_1$, we first define the marginal distribution as $X_1 \sim F_1(\boldsymbol{\Theta}_0)$. We then proceed by defining the conditional distribution of $X_2 | X_1 \sim F_2(\boldsymbol{\Theta}_1, X_1),$ and recursively defining the other distributions for the remaining covariates, $F_s$, for $s = 1, 2, \dots, S$. This nested regression framework is particularly useful when sufficient information is available to specify these conditional relationships. In practice, the conditioning set can be reduced whenever conditional independence assumptions are appropriate, so that each covariate is modeled only as a function of the variables that directly influence its distribution. When less information is available, alternative approaches such as copula-based methods \cite{nelsen2006introduction} or simulation-based reordering algorithms \cite{iman1982distribution} may also be used to induce realistic correlations among variables.

As our study is a hypothetical RCT, the exposure variable $Z$ can be randomly assigned, e.g., 
\[
Z_j \overset{iid}\sim  \text{Bernoulli}(q), \quad j=1,..,n,
\]
where $q$ is the expected proportion of treated individuals. Equivalently, the first $\lfloor nq \rfloor$ individuals can be assigned a $1$ (indicating treated), while the remaining $\lceil n(1-q) \rceil$ can be assigned $0$. Finally, for each individual, we define the distribution of the outcome given the covariates and the exposure variable as 
\[
Y | \boldsymbol{X}, Z \sim F(\boldsymbol{\Theta}, \boldsymbol{X}, Z),
\]
where $F$ depends on the covariates $\boldsymbol{X}$ and exposure variable $Z$ and a set of parameters for the data-generating process $\boldsymbol{\Theta}$. Importantly, note that to compute EVSI, the set of parameters $\boldsymbol{\Theta}$ must include at least one of the underlying health economic model's parameters $\boldsymbol{\theta}$, which will be the parameter(s) informed by future data collection.

In this way, we define the joint distribution of data from a hypothetical RCT,
$(\boldsymbol{Y}_{\mathrm{RCT}}, \boldsymbol{X}, \boldsymbol{Z})$, where
$\boldsymbol{Y}_{\mathrm{RCT}}$ is the vector of outcomes,
$\boldsymbol{X}$ is the $n \times S$ matrix of covariates, and
$\boldsymbol{Z}$ is the vector of treatment assignments.
A simulated dataset from this distribution is denoted by
$(\boldsymbol{y}_{\mathrm{RCT}}, \boldsymbol{x}, \boldsymbol{z})$.

We now describe how to transform this randomized dataset into observational data exhibiting confounding through population imbalance.

\subsubsection{Inducing confounding through stochastic population imbalance}

To generate observational data from the simulated hypothetical RCT data, we induce systematic differences in the covariate distribution between treated and control groups. We assume that one or more covariates $X^{(i)}$ influence treatment assignment, thereby creating population imbalance between the two groups. For example, older individuals may be more likely to receive treatment than younger individuals, resulting in treated and control populations that differ with respect to age. The magnitude of the induced imbalance is specified by the investigator to reflect plausible characteristics of the observational study being planned and is implemented through the stochastic selection mechanism described below.

To induce bias in our simulated RCT dataset $D_{RCT}$, we introduce a probability $p(x^{(i)})$ that defines the probability that an individual with covariate value $x^{(i)}$ is discarded from the dataset. The probability is specified so that:
\[
p\left(x_s^{(i)}\right) > p\left(x_t^{(i)}\right) 
\quad \text{if} \quad x_s^{(i)} > x_t^{(i)} \quad \text{and} \quad Z_s = Z_t = 1,
\]
\[
p\left(x_s^{(i)}\right) > p\left(x_t^{(i)}\right) 
\quad \text{if} \quad x_s^{(i)} < x_t^{(i)} \quad \text{and} \quad Z_s = Z_t = 0,
\]
which means that higher values of the selected covariate decrease the probability of remaining in the treated group and increase the probability of remaining in the control group. The covariate driving the imbalance, the direction of this relationship, and the functional form of $p(x^{(i)})$ should be specified to reflect the expected treatment allocation mechanism in the clinical setting of interest. These choices may be informed by substantive clinical expertise, previous observational studies, or expert elicitation. When substantial uncertainty exists regarding the expected degree of confounding, the proposed framework can also be applied across a range of plausible imbalance scenarios as part of a sensitivity analysis.

\noindent For each individual $j$, we sample an indicator
\[
\pi_j \sim \text{Bernoulli}(p(x_j^{(i)})),
\]
where $\pi_j = 1$ indicates that the individual is removed from the dataset. Thus, individuals with $\pi_j = 1$ are discarded from $D_{RCT}$, which can then be repeated until a pre-specified number $N_{dis}$ of individuals has been removed from each treatment group.

The resulting dataset, denoted $D_{obs}$, exhibits systematic differences in the distribution of $X^{(i)}$ between treated and control populations. Importantly, outcomes are not altered but as they are related to $X^{(i)}$, confounding arises as the treatment groups now differ with respect to baseline characteristics.

The strength of confounding can be controlled by modifying the functional form of $p(x^{(i)})$ and varying $N_{dis}$. The more people discarded from the RCT data, the greater the population imbalance. Note that this method can be extended to scenarios that combine multiple covariates to induce group differences by applying dimensionality reduction techniques such as principal component analysis (PCA) \cite{pearson1901} or factor analysis \cite{spearman1904}, or by discarding individuals located in extreme regions of $\mathbb{R}^n$ rather than $\mathbb{R}$. Other methods for inducing confounding under different assumptions have also been proposed in the literature \cite{autore2026titolo}, providing alternative frameworks when different structural or informational constraints apply. The full procedure for transforming the randomized dataset into observational data is summarised in Table \ref{box:decoupling}.

To quantify the degree of induced confounding, we use the population Mahalanobis distance \cite{mclachlan1999mahalanobis} between the treatment and control groups with respect to the confounding covariates. Since the selection mechanism induces confounding through systematic imbalance in these covariates, the Mahalanobis distance provides a convenient summary measure of the resulting multivariate imbalance. 

    \[
    \Delta_{N_{dis}} = \sqrt{(\boldsymbol{\mu}_t-\boldsymbol{\mu}_c)^T\left(\frac{\boldsymbol{S}_c + \boldsymbol{S}_t}{2}\right)^{-1}(\boldsymbol{\mu}_t-\boldsymbol{\mu}_c)}
    \]
The Mahalanobis distance between the mean values $\mu_t,\mu_c$ of the treated and control populations with respect to the inner product $\left(\left(\boldsymbol{S}_c + \boldsymbol{S}_t\right)/2\right)^{-1}$, where $\boldsymbol{S}_t, \boldsymbol{S}_c$ are the sample covariates matrix, respectively, of the treated and control individuals; and $\mu_t$ and $\mu_c$ represent the mean of the covariates distributions for the treatment and control group respectively.
The notation $\Delta_{N_{dis}}$ indicates that the Mahalanobis distance is computed after removing $N_{dis}$ individuals according to the selection mechanism. In general, larger values of $N_{dis}$ lead to greater induced covariate imbalance, so $\Delta_{N_{dis}}$ is used to denote the imbalance associated with a given value of $N_{dis}$. Note that, alternative measures of covariate imbalance, such as standardized mean differences or propensity score diagnostics \cite{stuart2010matching}, could also be used within the proposed framework.

\begin{table}
\begin{tcolorbox}
\textbf{Method to Induce Bias in Simulated Data}
\begin{enumerate}
    \item Start with a synthesized RCT $D_{RCT}$ and select a particular covariate $\boldsymbol{X}^{(i)}$.
    \item Define the probability $p(x^{(i)})$ of an individual with covariate profile $x^{(i)}$ to be discarded so that 
    \[
    p(x_s^{(i)}) > p(x_t^{(i)}) \quad \text{if} \quad x_s^{(i)} > x_t^{(i)} \quad \text{and} \quad Z_s = Z_t = 1
    \]
    \[
    p(x_s^{(i)}) > p(x_t^{(i)}) \quad \text{if} \quad x_s^{(i)} < x_t^{(i)} \quad \text{and} \quad Z_s = Z_t = 0
    \]
   Note that, without loss of generality, we assume this direction of the relationship between the covariate values and the probability of inclusion; the opposite direction could equivalently be considered. Moreover, 
    \[
    p(max(X_s^{(i)})) = p(min(X_t^{(i)})) = 1 \quad for \quad Z_s = 1, Z_t = 0
    \]
     $p(x^{(i)})$ represents the probability of an individual with covariate $X^{(i)} = x^{(i)}$ to be discarded from the data;

    \item For each individual sample $\pi_j = Bern(p(x_j^{(i)}))$, the missing indicator ($\pi_j = 1$ if individual $j$ misses and $\pi_j = 0$ otherwise).
    
    \item Select individuals $\{I_1,.., I_k\}$ from $D_{RCT}$ such that $\pi_{I_i} = 1$ (missing indicator for individual $I_i$) and $Z_{I_i}=1$ for $i=1,..,k$, and $h$ individuals $\{J_1,.., J_h\}$ from $D_{RCT}$ such that $\pi_{J_i} = 1$ and $Z_{J_i}=0$ for $i=1,..,h$.
    \item Remove the selected individuals from $D_{RCT}$.
    \item Repeat until the prespecified number $N_{dis}$ of individuals to discard from each population is reached and $D_{obs}$ is obtained.
\end{enumerate}
\end{tcolorbox}
\caption{Method to Induce Bias in Simulated Data}
\label{box:decoupling}
\end{table}

\subsection{Calculating EVSI accounting for confounding}
\label{EVSI_OBSERVATIONAL}

In the previous section, we defined the general data-generating mechanism for future observational data, where the outcome distribution depends on parameter(s) from the health economic model, $\boldsymbol{\theta}$, that would be informed by the future study. This section discusses how this data-generating mechanism should be incorporated into EVSI calculation using the regression-based method.

\subsubsection{Simulating the covariates of interest}
\label{evsi_obs}

To compute EVSI when future data are observational and affected by confounding, we first simulate the covariates for each individual $\boldsymbol{x}_j$ and the exposure variable $z_j$, $j =1, \dots, n$, following the nested regression structure from above. Note that these covariates are fixed throughout the EVSI calculation. Therefore, for each probabilistic analysis (PA) simulation $\boldsymbol{\theta}^{(i)}$, $i = 1, \dots, K$, the simulated datasets differ only in the outcome realizations, while sharing the same covariate profile. EVSI is thus first computed conditional on the simulated covariate profile. Finally, to obtain the unconditional EVSI, we integrate over the distribution of the covariates. In practice, this requires repeating the EVSI calculation $M$ times using independently simulated covariate profiles and averaging the resulting estimates. In practice, $M$ may be selected by increasing the number of simulated covariate profiles until the estimated EVSI stabilizes. The appropriate value of $M$ is application-specific and depends on the complexity of the underlying health economic model and the covariate-generating mechanism.

\subsubsection{Simulating the patient-level outcomes}

Once the covariates have been simulated and fixed for the EVSI calculation, we must generate the complete RCT outcomes, conditional on these covariates $\boldsymbol{x}$ and the exposure $\boldsymbol{z}$. Recall that the data-generating model for the outcomes must depend on the PA simulations $\boldsymbol{\theta}^{(i)}$, $i = 1, \dots, K$. Following the general framework introduced in the previous section, outcomes are generated from the conditional distribution $Y \mid \boldsymbol{X}, Z \sim F(\boldsymbol{\Theta}, \boldsymbol{X}, Z)$, where $\boldsymbol{\Theta}$ includes the health economic parameter(s) $\boldsymbol{\theta}$ informed by the future study. To implement this distribution within the EVSI calculation, we define individual-level parameters that incorporate the effect of the simulated covariates.
\[
\tilde{\boldsymbol{\theta}}^{(i)} = f(\boldsymbol{\theta}^{(i)}, \boldsymbol{X}),
\]
where $f$ represents the dependence structure between the health economic parameter(s) $\boldsymbol{\theta}$ and the individual covariates. The individual-level parameters $\tilde{\boldsymbol{\theta}}^{(i)}$ define the components of $\boldsymbol{\Theta}$ that vary across individuals in the conditional outcome distribution $F(\boldsymbol{\Theta}, \boldsymbol{X}, Z)$. For instance, a linear specification may be adopted:
\[
\tilde{\boldsymbol{\theta}}^{(i)} 
=
\boldsymbol{\theta}^{(i)} 
+
\sum_{s=1}^{S} \beta_s (X_s - \overline{X})
\]
The linear specification is adopted here for simplicity and ease of interpretation. More generally, the function $f$ may take any form that adequately represents the relationship between the individual covariates and the health economic parameter(s). More flexible specifications, such as nonlinear functions or interaction terms, may be appropriate when supported by substantive knowledge or empirical evidence regarding the data-generating process.

For each individual $j$, outcomes are then simulated from
$y_j^{(i)}\sim F(\boldsymbol{\Theta}^{(i)},\boldsymbol{x}_j,z_j)$, where $\boldsymbol{\Theta}^{(i)}$ incorporates the individual-level parameters $\tilde{\boldsymbol{\theta}}^{(i)}_j$ corresponding to the health economic parameter(s). This yields complete simulated RCT datasets 
$(\boldsymbol{y}^{(i)}_{RCT}, \boldsymbol{x}, \boldsymbol{z})$,
for $i = 1, \dots, K$. These datasets correspond to the complete RCT datasets defined in the previous section and represent the starting point for generating the observational datasets used in the EVSI calculation.

\subsubsection{Generating observational data with confounding}

Following the procedure described in the previous section, we transform each complete simulated RCT dataset into a synthetic observational dataset by inducing confounding. Concretely, for each simulated dataset $(\boldsymbol{y}^{(i)}_{RCT}, \boldsymbol{x}, \boldsymbol{z})$, we randomly remove individuals from the dataset according to a probability function depending on a chosen covariate (or covariates) and the related treatment group, thereby inducing systematic confounding in the simulated data. The resulting dataset is denoted 
$(\boldsymbol{y}^{(i)}_{obs}, \boldsymbol{x}_{obs}, \boldsymbol{z}_{obs})$.

Because treatment assignment is now associated with baseline covariates, the naive summary statistic $T(\boldsymbol{y}^{(i)}_{obs})$ does not generally give an unbiased estimate of the same statistic computed on the original individual-level randomized data. Therefore, we will apply inverse probability weighting (IPW) assigning a weight to each individual in the data to obtain an unbiased estimate of the relevant estimand. The weights assigned to each individual are $w_i = 1/e(\boldsymbol{\tilde{X}})$ if the individual $i$ belongs to the treatment group, and $w_i = 1/(1-e(\boldsymbol{\tilde{X}}))$ if the individual $i$ belongs to the control group. Here, $e(\boldsymbol{\tilde{X}})$ is the propensity score, defined as $e(\boldsymbol{\tilde{X}}) = pr(Z = 1 | \boldsymbol{\tilde{X}})$, i.e. the probability of being treated given the individual's confounders' profile $\boldsymbol{\tilde{X}}$ \cite{hernan2020causal} \cite{austin2015moving}. Note that, in general, confounders $\boldsymbol{\tilde{X}} \subseteq \boldsymbol{X}$ are the subset of the covariates that are directly associated with both the outcome and the exposure variable. In this way, we obtain a weighted estimator $\tilde{T}(\boldsymbol{y}^{(i)}_{obs})$, so that:
\[
\mathbb{E}\left[\tilde{T}(\boldsymbol{y}^{(i)}_{obs})\right]
=
\mathbb{E}\left[T(\boldsymbol{Y}^{(i)}_{RCT})\right].
\]

\subsubsection{Computing EVSI with observational data}

For each covariate profile, EVSI is computed using the regression-based approach. Specifically, we fit a nonparametric regression between $\text{NB}(\boldsymbol{\theta}^{(i)})$ and $\tilde{T}(\boldsymbol{y}^{(i)}_{obs})$. The fitted values $\hat{g}_d(\tilde{T}(\boldsymbol{y}^{(i)}_{obs}))$ are used to estimate

\begin{equation}
\label{eq:evsi_conf}
\text{EVSI} \mid \boldsymbol{x}
=
\frac{1}{K}\sum_{i = 1}^K 
\max_d \hat{g}_d(\tilde{T}(\boldsymbol{y}^{(i)}_{obs}))
-
\max_d
\frac{1}{K}\sum_{i = 1}^K 
\hat{g}_d(\tilde{T}(\boldsymbol{y}^{(i)}_{obs})).
\end{equation}

Finally, we integrate out the dependence on $\boldsymbol{X}$ using Monte Carlo computation:

\[
\text{EVSI}
=
\mathbb{E}_{\boldsymbol{X}}(\text{EVSI} \mid \boldsymbol{X})
\approx
\frac{1}{M}
\sum_{m=1}^{M}
\text{EVSI} \mid \boldsymbol{x}_m.
\]

This process is summarised in Box \ref{box:evsi_obs}.

\begin{table}[htbp]
\centering
\begin{tcolorbox}
\textbf{EVSI with observational data affected by confounding}
\begin{enumerate}

    \item Generate covariates $\boldsymbol{X}$ and exposure variables $\boldsymbol{Z}$ for $n$ individuals.
    
    \item Sample $K$ parameter sets $\boldsymbol{\theta}^{(i)}$ from the probabilistic analysis.
    
    \item For each $\boldsymbol{\theta}^{(i)}$, construct individual-level parameters 
    $\tilde{\boldsymbol{\theta}}^{(i)}$ and simulate complete RCT outcomes
    \[
    \boldsymbol{Y}_{RCT}^{(i)} \sim F(\tilde{\boldsymbol{\theta}}^{(i)}).
    \]
    
    \item Transform each simulated RCT dataset into observational data 
    $(\boldsymbol{Y}_{obs}^{(i)}, \boldsymbol{X}_{obs}, \boldsymbol{Z}_{obs})$
    by inducing stochastic population imbalance.
    
    \item Apply inverse probability weighting to obtain the adjusted summary statistic
    $\tilde{T}(\boldsymbol{Y}_{obs}^{(i)})$, targeting the corresponding randomized estimand.
    
    \item Fit nonparametric regression models
    \[
    \text{NB}_d(\boldsymbol{\theta}^{(i)}) = g_d\left(\tilde{T}(\boldsymbol{Y}_{obs}^{(i)})\right) + \epsilon
    \]
    to estimate the preposterior mean.
    
    \item Compute EVSI conditional on the sampled covariate profile.
    
    \item Integrate over covariate profiles to obtain the final EVSI estimate.

\end{enumerate}
\end{tcolorbox}
\caption{Simulation procedure for computing EVSI with observational data affected by confounding}
\label{box:evsi_obs}
\end{table}

\subsubsection{Adjusting sample size to account for confounding}

We will show that, as expected, EVSI computed using observational data affected by confounding is lower than EVSI computed using the complete randomized dataset. This reduction arises because confounding and the subsequent weighting procedure introduce additional layers of variability in the estimator of the summary statistic. In particular, the additional variance arises from two sources: first, because the stochasticity of the method we use to induce confounding means we do not know the exact level of bias in advance; second, from the IPW method used to adjust confounding itself. Notice that, by inducing bias by discarding individuals, the sample size of the resulting observational dataset decreases by $2N_{dis}$. For this reason, to obtain comparable EVSI profiles, we start from a sample size of $\tilde{n} = n + 2N_{dis}$ in the observational EVSI calculation. This adjustment is introduced solely to ensure that, after the selection procedure, the observational and randomized datasets have the same final sample size, allowing a fair comparison of their EVSI profiles.

In many study design scenarios, we are interested in determining the sample size of observational data required to achieve the same value of information as a randomized study of size $n$. In other words, we seek the observational sample size that compensates for the loss of precision induced by confounding and adjustment.

A first idea would be to adjust the sample size by computing the average effective sample size from the IPW weights obtained across the different simulated observational data. Under standard assumptions, it is usually assumed that the design effect $\text{DEFF} = n/\text{ESS}$ is constant, implying that the distribution of the inverse probability weights remains stable as $n$ increases. In this scenario, $n_{\text{ESS}} = n^2/\overline{\text{ESS}}$ is assumed to be linear in $n$, where $\overline{\text{ESS}}$ is the average effective sample size. Nonetheless, this adjustment does not generally accurately recover the original level of EVSI. This is because standard assumptions rely on $n_{\text{ESS}}$ being linear in $n$, while $\text{EVSI}(n)$ is generally a concave function of $n$. Consequently, applying a linear correction to the sample size introduces systematic bias when attempting to match the original information level. Depending on the curvature of the \text{EVSI} function, this bias may manifest as underestimation or overestimation of the required sample size. In the Results section, we will show an example in which the EVSI is underestimated.

Since we know that the larger the sample size of simulated data, the higher the EVSI \cite{heath2019estimating}, a straightforward approach would be to sequentially increase the observational sample size and recompute EVSI until the value matches that obtained under randomization. However, this approach is computationally intensive, as it requires repeatedly performing the full nonparametric regression procedure.

For this reason, we rely on a computationally efficient method to determine the required sample size adjustment recently introduced in the context of EVSI computation with missing data \cite{benetti2026calculatingexpectedvaluesample}. The key idea is to identify a sample size such that the variance of the adjusted summary statistic computed from observational data, 
$\tilde{T}(\boldsymbol{Y}_{obs})$, matches the variance of the summary statistic computed from the complete randomized data, $T(\boldsymbol{Y}_{RCT})$. This approach requires simulating observational datasets of increasing size and computing the weighted estimator $\tilde{T}(\boldsymbol{Y}_{obs})$, but avoids repeatedly fitting the nonparametric regression model and recomputing EVSI. The identified sample size should then be validated by recomputing EVSI and verifying that the target EVSI level is recovered.

Because $\tilde{T}(\boldsymbol{Y}_{obs})$ is constructed to be unbiased for $T(\boldsymbol{Y}_{RCT})$, matching the first and second moments of the summary statistic provides an approximation to the information content of the randomized study. Consequently, the resulting EVSI is expected to closely match the EVSI obtained under randomization \cite{heath2018efficient}.

\section{Application}

In this section, we describe two models used to compute EVSI with observational data affected by confounding. The first assumes a Normal–Normal model for computational simplicity, while the second uses a realistic underlying health-economic model combining a decision tree and a Markov state transition model \cite{heath2024value}.

\subsection{Normal–Normal model}
\label{evsi_obs_normal}

For the probabilistic analysis, we assume prior information on the mean outcome in the treatment and control groups. Both follow Normal distributions with means $\theta_1 = 40$ and $\theta_0 = 20$, and variances $\sigma_1^2 = 8$ and $\sigma_0^2 = 6$, respectively. Formally,

\[
\theta_{Z=0}^{(i)} \sim N(\theta_0,\sigma_0^2), 
\quad 
\theta_{Z=1}^{(i)} \sim N(\theta_1,\sigma_1^2),
\quad i=1,\dots,K.
\]

The proposed study collects data from $n = 100$ individuals with $n_0 = n_1 = 50$. Outcomes are normally distributed with variance $\sigma_Y^2$, and individual means depend on covariates.

The net benefit functions are

\begin{equation}
\begin{aligned}
\text{NB}_0(\theta) &= k\theta_0, \\
\text{NB}_1(\theta) &= k\theta_1 - c.
\end{aligned}
\end{equation}
Each individual has two covariates: age and number of comorbidities $(age, com)$. These are generated as

\[
age_j \sim \mathcal{N}(\mu_{age},\sigma^2_{age}),
\quad
com_j \sim Bin(n_{com},p_{com}),
\]
with $\mu_{age}=57$, $\sigma_{age}=8$, $n_{com}=10$, and $p_{com}=0.5$. For each PA simulation, individual-level means are

\[
\tilde{\theta}_j^{(i)} 
=
\theta_{Z_j}^{(i)}
+
\alpha_1(age_j-\overline{age})
+
\alpha_2(com_j-\overline{com}),
\]
with $\alpha_1 = 3$ and $\alpha_2 = 5$. Outcomes are generated as

\[
Y_j^{(i)} \sim N(\tilde{\theta}_j^{(i)}, \sigma_Y^2),
\]
yielding complete simulated RCTs $(\boldsymbol{y}^{(i)}_{RCT},\boldsymbol{x},\boldsymbol{z})$ for $i=1,\dots,K$ with $K=5000$.

To generate observational data, we induce population imbalance using the procedure described previously. In this example, age is selected as the covariate driving imbalance. Individuals with higher age are less likely to remain in the treated group and more likely to remain in the control group. This generates systematic differences in baseline characteristics between treatment arms. Applying this procedure to each simulated RCT produces observational datasets 
$(\boldsymbol{y}^{(i)}_{obs}, \boldsymbol{x}_{obs}, \boldsymbol{z}_{obs})$.

The naive estimator of the treatment effect computed from $\boldsymbol{y}_{obs}$ does not target the randomized estimand. Therefore, we estimate the propensity score model $\Pr(Z=1 \mid \boldsymbol{X})$, and compute inverse probability weights. The summary statistic $T$ is then defined as the weighted difference in mean outcomes between treatment groups, $\tilde{T}(\boldsymbol{y}_{obs})$, which consistently estimates the randomized treatment effect under correct specification. Finally, we fit a Generalized Additive Model (GAM) to estimate $\text{EVSI} \mid \boldsymbol{X}$ before integrating over $\boldsymbol{X}$.

\subsection{Chemotherapy treatment model}
\label{evsi_obs_markov}

The second application considers a health-economic model combining a decision tree and a Markov state transition model \cite{heath2024value}. The decision maker chooses between two chemotherapy treatments ($D=2$). Treatments differ in cost and side-effect profile, but not in efficacy. The model estimates the probability of experiencing side effects $s_d$ under standard care ($d=0$) and novel treatment ($d=1$). The Markov component models long-term health consequences and associated costs. The future observational study collects data on side effects. We consider a sample size of $n=1000$ with $n_0=n_1=500$ before imbalance is induced. Covariates are generated as in the Normal–Normal example. Individual probabilities of side effects are defined as
\[
\tilde{s}^{(i)}_{Z_j,j}
=
\Phi\left(
\Phi^{-1}(s_{Z_j}^{(i)})
+
\alpha_1(age_j-\overline{age})
+
\alpha_2(com_j-\overline{com})
+
\alpha_{0,Z_j}
\right),
\]
with $\alpha_1 = 0.05$ and $\alpha_2 = 0.3$. Outcomes are simulated as
\[
Y_j^{(i)} \sim \text{Bernoulli}(\tilde{s}_j^{(i)}),
\]
yielding complete RCT datasets.

Population imbalance is induced using the same method in \ref{box:decoupling}. Age is again selected as the driver of imbalance. The resulting observational datasets exhibit reduced overlap in covariate distributions across treatment groups. To adjust for confounding, we estimate propensity scores and compute inverse probability weights. The summary statistic $T$ is defined as the weighted log-odds ratio of side effects between treatment groups, $\tilde{T}(\boldsymbol{y}_{obs})$, which targets the randomized estimand.

\subsection{Comparison}
\label{comparison_obs}

For both examples, we use $K=5000$ PA simulations and $M=10$ covariate profiles. In the case of Normal-Normal model, we compare the following quantities:

\begin{enumerate}
    \item EVSI computed on the simulated RCT data with no induced imbalance (Original EVSI).
    
    \item EVSI computed on observational data after adjustment using inverse probability weighting.
    
    \item EVSI computed on observational data with adjusted sample size determined through variance matching.
\end{enumerate}

Notice again that, by applying our method, the sample size of the resulting observational dataset decreases by $2N_{dis}$; therefore, to obtain comparable EVSI profiles, we start from a sample size of $\tilde{n} = n + 2N_{dis}$.

Moreover, for the Chemotherapy treatment model, we will compare the above quantities by varying the number of individuals discarded, $N_{dis}$, thereby obtaining different levels of confounding $\Delta_{N_{dis}}$, measured as the Mahalanobis distance between the treatment and control groups. Finally, for the chemotherapy treatment model, we also show that adjusting the sample size by computing the average effective sample size across the different weights in the IPW procedure still does not reach the original EVSI level, as discussed in the previous section.

EVSI is plotted across a range of willingness-to-pay thresholds $k$. We summarise differences by computing the average absolute and relative differences between the Original EVSI and the observational EVSI estimates.

\section{Results}

\subsection{Normal–Normal model}

Figure 1 displays the Original EVSI computed from the simulated randomized data, the EVSI computed from observational data after inverse probability weighting (IPW) adjustment, and the EVSI computed from observational data with adjusted sample size.

\begin{figure}
    \centering
    \includegraphics[width=0.7\linewidth]{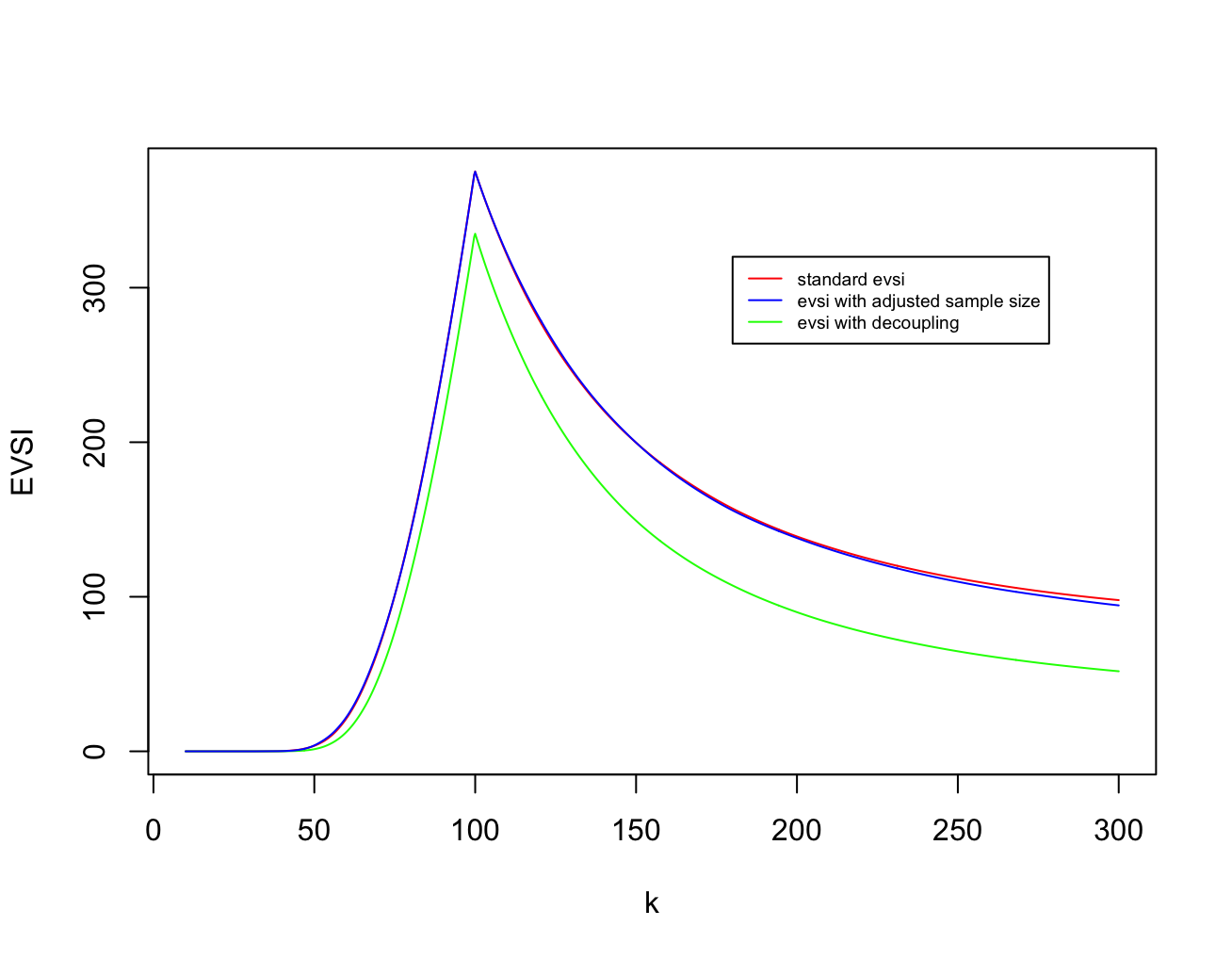}
    \caption{EVSI calculated using complete randomized data $(\boldsymbol{Y}_{\text{RCT}}^{(i)},\boldsymbol{X}, \boldsymbol{Z})$ in red, EVSI computed on IPW-adjusted observational data in green, and EVSI computed on IPW-adjusted observational data with adjusted sample size in blue. Normal-Normal model.}
\end{figure}

As expected, EVSI decreases when computed using observational data affected by confounding. This reduction is driven by the induced population imbalance, which reduces effective overlap between treatment groups and inflates the variance of the treatment effect estimator.

Figure 2(a) is used to determine the sample size required for the observational study to achieve the same value of information as the complete randomized study with $n=100$. The horizontal line represents the variance of $T(\boldsymbol{Y}_{RCT})$, while the dashed lines represent a tolerance band around this value.

\begin{figure}[h]
    \centering
\begin{subfigure}{0.48\textwidth}
        \centering
        \includegraphics[width=\linewidth]{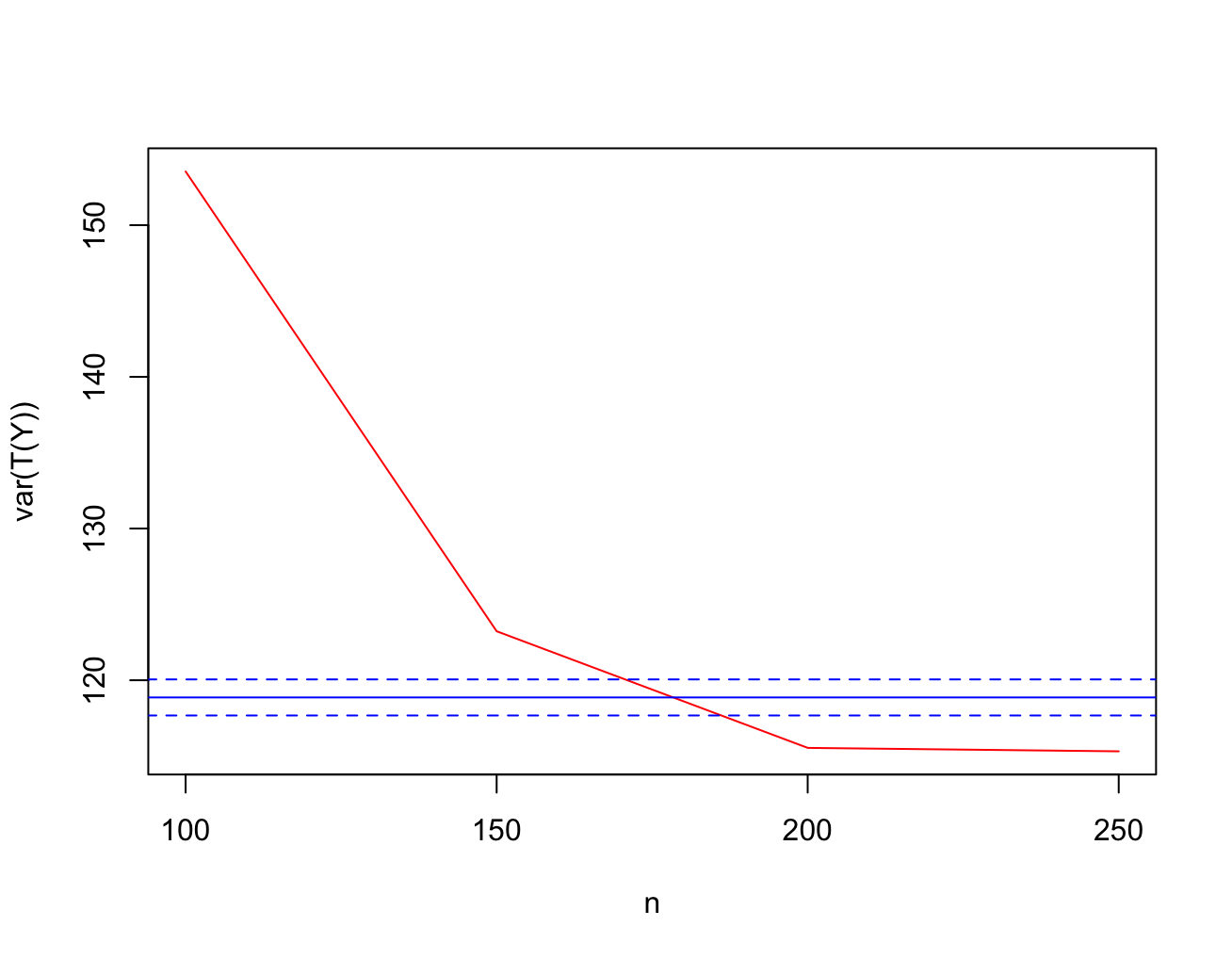}
        \caption{}
        \label{fig:a}
    \end{subfigure}
    \hfill
\begin{subfigure}{0.48\textwidth}
        \centering
        \includegraphics[width=\linewidth]{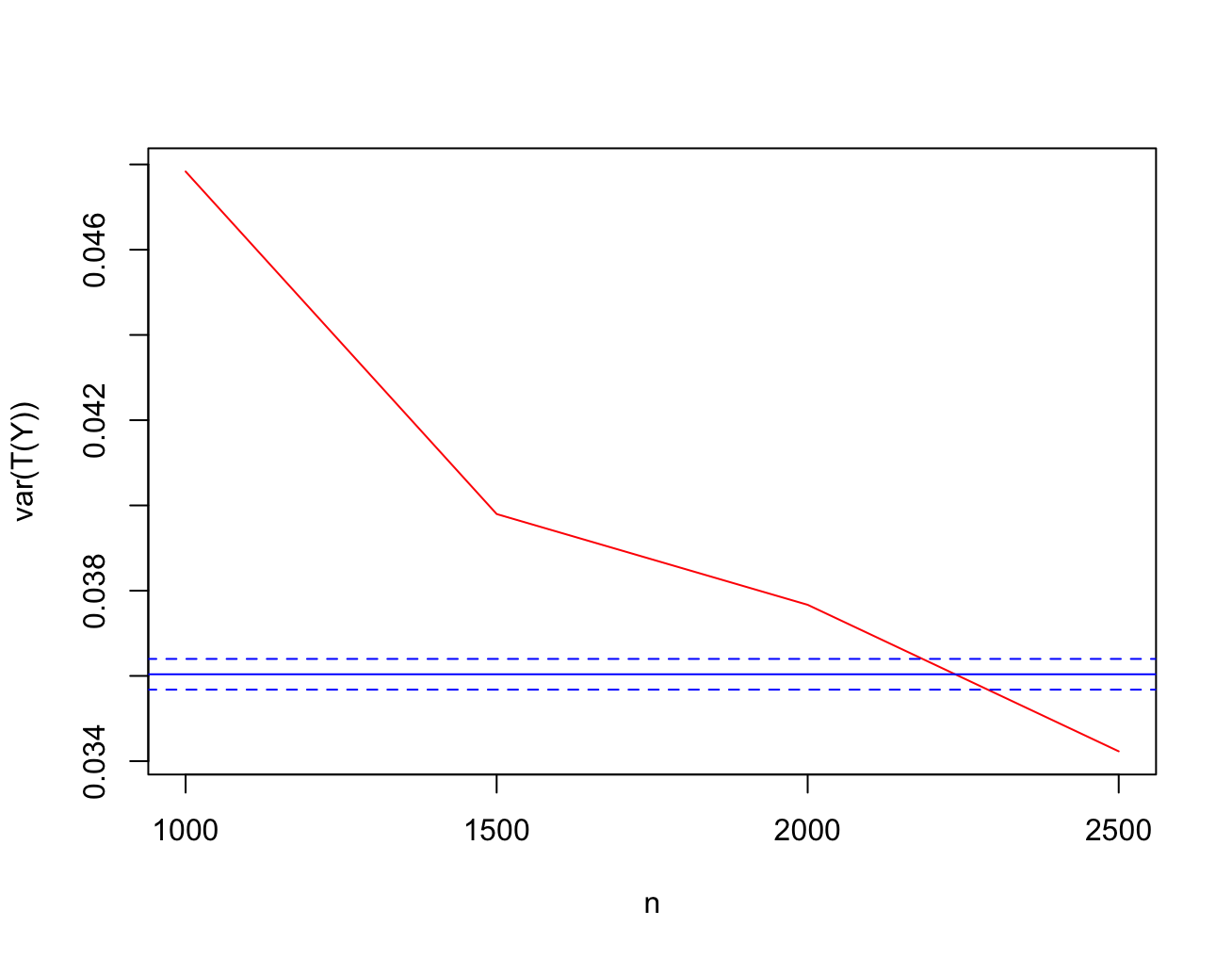}
        \caption{}
        \label{fig:b}
    \end{subfigure}
    
    \caption{Comparison between var($\tilde{T}(\boldsymbol{Y}_{obs})$) for increasing sample size $n$ and var($T(\boldsymbol{Y}_{RCT})$): (a) Normal–Normal model, (b) Chemotherapy treatment model.}
    \label{fig:comparison}
\end{figure}

From Figure \ref{fig:a}, we identify that $n_{adj} \approx 180$ produces a variance of the weighted summary statistic comparable to that of the randomized estimator. Thus, the observational sample size must increase from $100$ to approximately $180$ to achieve similar information content. 

When EVSI is recomputed using this adjusted sample size, the resulting curve closely matches the Original EVSI, confirming that variance matching provides an accurate and computationally efficient sample size correction.

\subsection{Chemotherapy treatment model}

\begin{figure}
    \centering
    \includegraphics[width=0.7\linewidth]{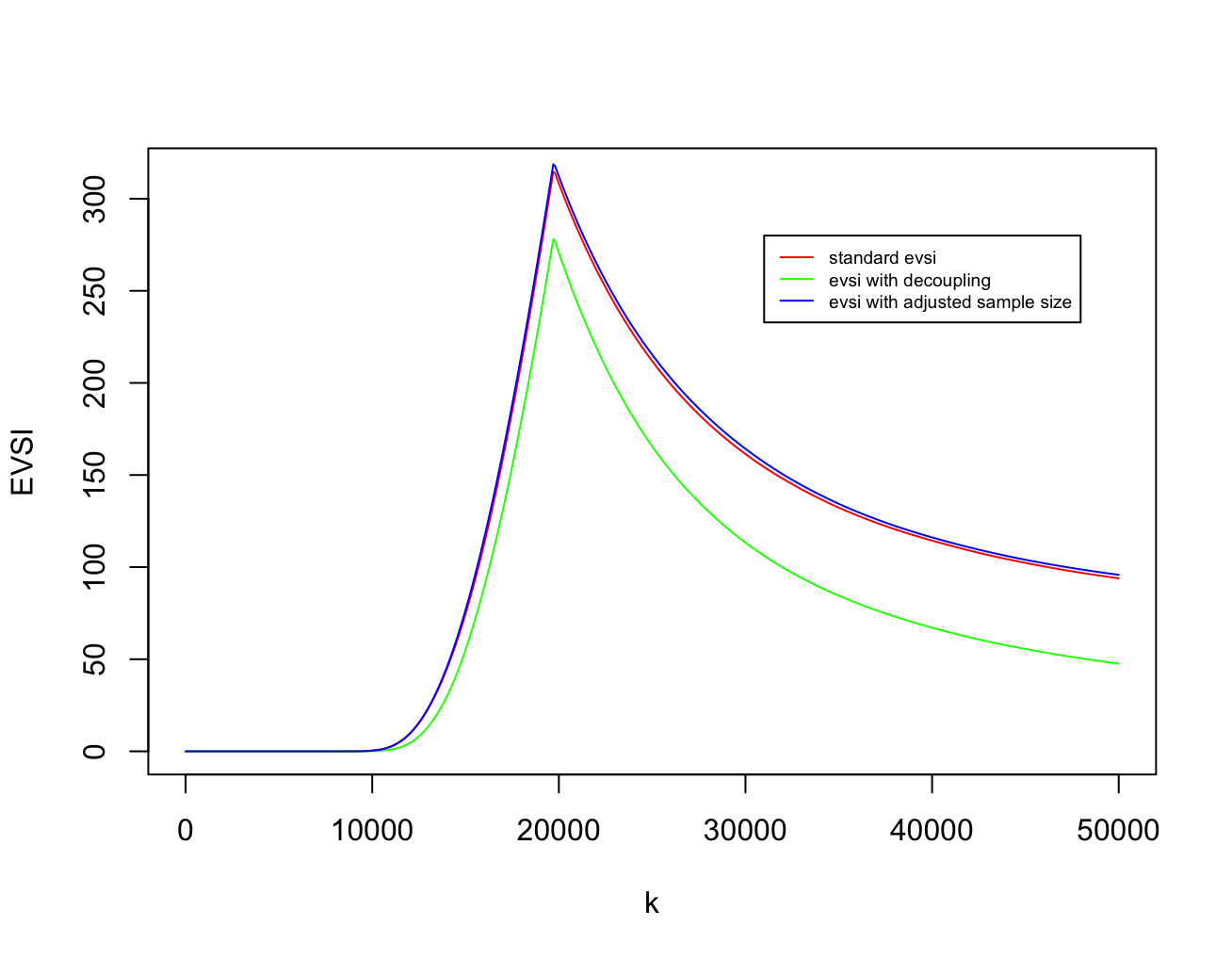}
    \caption{EVSI calculated using complete randomized data in red, EVSI computed on IPW-adjusted observational data in green, and EVSI computed on IPW-adjusted observational data with adjusted sample size in blue. Chemotherapy treatment model.}
    \label{a}
\end{figure}

Figure \ref{a} displays the corresponding EVSI curves for the chemotherapy model. The same pattern observed in the Normal–Normal example emerges in this more complex setting.

The EVSI computed from observational data is consistently reduced, even after IPW adjustment, reflecting loss of precision due to confounding-induced population imbalance.

To further investigate the impact of confounding, we progressively increase the degree of induced population imbalance by removing $N_{dis}=\{500,1000\}$ individuals, obtaining $\Delta_{500} = 1.149$ and $\Delta_{1000} = 1.21$. This allows us to explore how the value of information changes as the overlap between the treated and control groups decreases.

\begin{figure}
    \centering
    \includegraphics[width=0.7\linewidth]{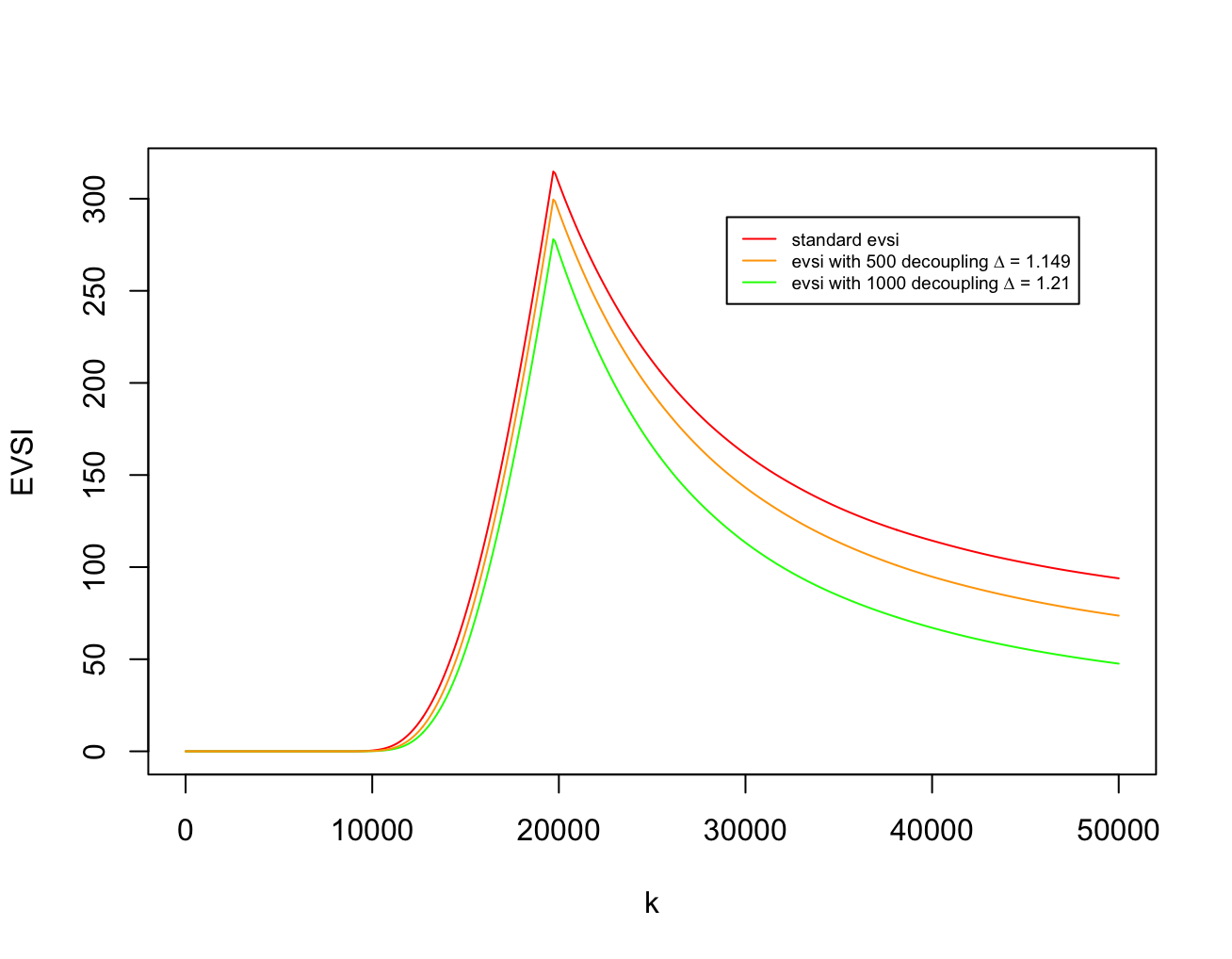}
    \caption{EVSI as a function of the degree of induced population imbalance in the chemotherapy treatment model. Higher imbalance corresponds to reduced covariate overlap between treatment groups.}
    \label{fig:bias_chemo}
\end{figure}

Figure \ref{fig:bias_chemo} shows a clear monotonic decline in EVSI as the degree of confounding increases. This behaviour reflects the growing variance of the weighted estimator and the associated reduction in effective sample size. As we already highlighted, the nominal sample size remains fixed across these scenarios; therefore, the observed decline in EVSI is entirely driven by the worsening overlap rather than by changes in recruitment size.

Figure 2(b) shows the variance-matching procedure for this model. The adjusted sample size required to match the variance of the randomized estimator is approximately $n_{adj} \approx 2250$, compared with the original randomized size of $n=1000$.

This substantial increase reflects the efficiency loss induced by weighting and limited overlap. After adjusting the observational sample size accordingly, the EVSI closely approximates the Original EVSI as shown in Figure \ref{a}, demonstrating that variance matching successfully restores the informational equivalence between observational and randomized designs.

\begin{figure}
    \centering
    \includegraphics[width=0.7\linewidth]{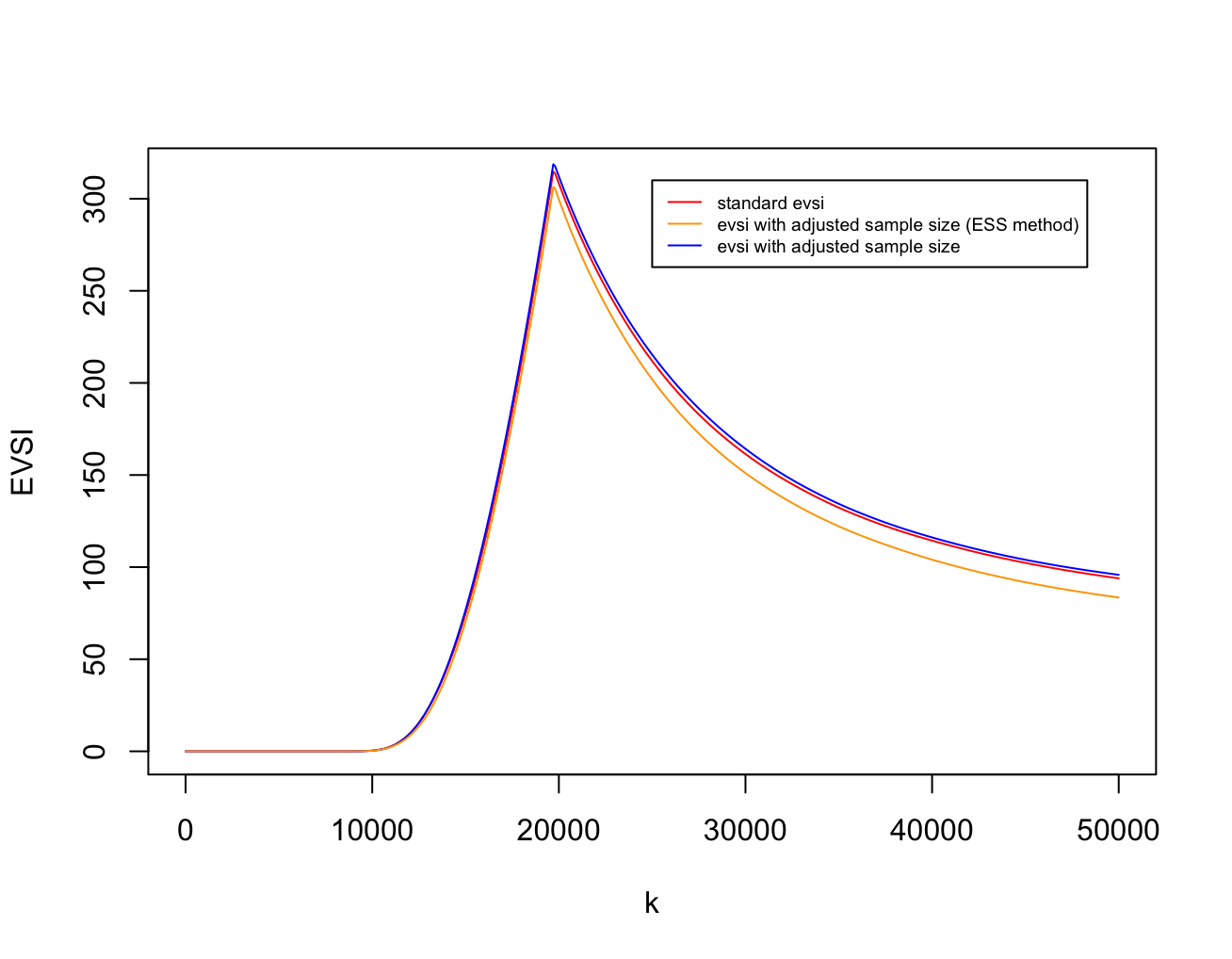}
    \caption{EVSI calculated using complete randomized data in red, EVSI computed on IPW-adjusted observational data with adjusted sample size with standard ESS method in orange, and EVSI computed on IPW-adjusted observational data with adjusted sample size in blue.}
    \label{b}
\end{figure}

Finally, Figure \ref{b} shows the underestimation of the required sample size that one would obtain using standard methods based on effective sample size calculations. In particular, in this experiment, we obtained $\overline{ESS} = 557.7$ and therefore $n_{ESS} = 1792$.

\section{Discussion}

In this paper, we introduced a methodology that incorporates realistic observational data into EVSI calculations. Specifically, we focused on settings in which future evidence is expected to arise from non-experimental studies affected by confounding. This extends the applicability of EVSI beyond idealized randomized designs and allows value-of-information analysis to be performed in contexts where randomized trials are infeasible or impractical.

When computing EVSI using observational data, we identified two key drivers of information loss. First, stochastic confounding-induced imbalance reduces covariate overlap when the true level of confounding is unknown a priori, which is entirely realistic in practice. Second, adjustment procedures such as inverse probability weighting inflate variability and reduce the effective sample size relative to the nominal sample size. Even when the adjustment procedure consistently targets the randomized estimand, the informational content of the observational study remains lower than that of a comparable randomized design.

Our results demonstrate that EVSI decreases as the degree of population imbalance increases. Importantly, this reduction occurs even when the nominal sample size remains constant, highlighting that the value of additional information depends critically on covariate overlap. In this sense, EVSI reflects not only the quantity of data collected but also the structural properties of the data-generating mechanism.

We also showed that standard sample size adjustment based on effective sample size calculation is insufficient to recover the value of information attainable under randomization. To address this issue, we integrated a computationally efficient variance-matching approach that identifies the observational sample size needed to replicate the variance of the randomized estimator, thereby restoring the original EVSI without repeatedly recomputing the full regression-based procedure.

Future research could explore the theoretical relationship between covariate overlap, effective sample size, and EVSI more formally. In particular, deriving analytical links between weighting-induced variance inflation and the loss in EVSI would allow for a deeper theoretical understanding of the economic value of randomization itself. Another promising direction would be to investigate alternative adjustment methods, such as matching or doubly robust estimators, and compare their impact on EVSI. This would provide a principled framework for evaluating causal inference methods based on their ability to preserve economic value of information within decision models.

\renewcommand{\bibname}{References}
\bibliographystyle{plain}
\bibliography{bib1.bib}

\end{document}